\documentclass[apj,numberedappendix]{emulateapj}
\usepackage{pxfonts}
\usepackage{color}

\shorttitle{Emission Spectra from Cyg X-3}
\shortauthors{Zhang \& Lu}

\begin{document}


\title{Broadband Emission Spectra from the Cygnus X-3 Jet in the Soft Spectral State}


\author{Jian-Fu Zhang\altaffilmark{1,2} and  Ju-Fu Lu\altaffilmark{1}}


\altaffiltext{1}{Department of Astronomy and Institute of Theoretical Physics and Astrophysics, Xiamen University, Xiamen, Fujian 361005, China;}
 \email{jianfuzhang.yn@gmail.com; lujf@xmu.edu.cn}  %
\altaffiltext{2}{Department of Physics, Tongren University, Tongren 554300, China}



\begin{abstract}
In order to understand the main observational characteristics of the Galactic X-ray binary Cygnus X-3, we propose a radiation model in which high-energy electrons accelerated in the dissipation zone of a jet produce the non-thermal broadband emissions. Broadband spectral energy distributions are computed to confront with the AGILE and\emph{ Fermi} LAT data together with the multi-band data, during soft X-ray spectral states. By fitting observations at different locations of the jet, we find that the emission region is rather compact and should be located at the distance of about one orbital radius. Our results can explain the current multi-frequency observations and also predict the TeV band emission.  The model could be tested by a polarization measurement at IR band, and/or by a correlation study between the GeV and TeV bands once very high-energy observations are available.
\end{abstract}



\section{INTRODUCTION}
In transient X-ray binaries, the hardness-intensity diagram (HID) is an important tool towards understanding the accretion disk/jet connection. The HID for a typical X-ray binary following a transient outbursting cycle exhibits a Q-type shape (\citealt{Fender04}), which consists of three canonical spectral states: low/hard (LH), high/soft (HS), and very high/intermediate states (VHS/IS). More details may refer to \cite{Remillard06}, \cite{Belloni10}, and \cite{Fender12}. This HID suggests two types of jets in the transient X-ray binaries. One steady, continuous jet, which has distributed dissipation along the jet, is in the LH state with an evident characteristic of the flat radio spectra. The other transient jet, which has an optically thin radio spectrum and high levels of polarization, appears in the VHS/IS with high-energy tails. It should be noted that in this canonical HID no jet appears in the HS state (see also Fig. 7 of \citealt{Fender04}).

Cygnus X-3 (Cyg X-3) was first discovered in the X-rays by \cite{Giacconi67}. The X-ray spectra of Cyg X-3 are complex, and the X-ray emission fluxes are modulated with the orbital period of the system. This source shows recurrent activities of relativistic jets, and is one of the brightest Galactic transient radio sources (\citealt{GK72}). It is shown that correlations between the hard X-ray flux and radio flux are switched from an anti-correlation to a positive correlation during the period of a radio outburst. In addition, the hard X-ray fluxes are always anti-correlated with soft X-rays (\citealt{McCollough99}). It seems that the HID for Cyg X-3 shows the canonical X-ray spectral states presented above. However, there are some significant differences. First, the flaring data fill the entire HID space, that is, Cyg X-3 exhibits a ``shoe" shape rather than a Q-type shape seen in a typical X-ray binary (\citealt{Weng13}). This strange shape of the HID may be due to very strong X-ray absorption in this source (e.g., \citealt{Szostek08}). Second, this source does not display hysteresis in the HID (\citealt{Hjalmarsdotter09}), which seems to require the source to be a transient. Persistent black hole X-ray binaries, such as Cygnus X-1, indeed do not show hysteresis (\citealt{Zdziarski04}). Third, the LH state for Cyg X-3 is only constrained in the ``toe" regime. However, the LH state for a typical source corresponds to the entire vertical branch (on the right side) of the HID.

Due to the above differences in the HID of Cyg X-3, some studies have been inspired to propose its state definition based on the X-ray and radio observations (\citealt{Waltman96,McCollough99,Szostek08,Hjalmarsdotter09,Koljonen10}). Comparisons between different classification methods may refer to Table 1 of \cite{Koljonen10}. The X-ray states proposed by \cite{Koljonen10} are shown as follows: quiescent, transition, flaring hard X-ray (FHXR), flaring intermediate (FIM), flaring soft X-ray (FSXR), and hypersoft states. The hypersoft state being associated with the quenched radio state, which is even softer than the ultrasoft state, presents a high-energy tail. In this state, radio fluxes fall to very low levels or no radio signature is seen. We note that in this classification there is a ``jet line" between the hypersoft and FSXR states, which is different from that of the canonical HID (\citealt{Koljonen10}). In particular, except the quiescent, transition, and FHXR states, the jet of Cyg X-3 also appears in both the FSXR and FIM states, which correspond to the canonical HS state (see Table 1 of \citealt{Koljonen10}). It seems that the jet in Cyg X-3 may be transient.

The $\gamma$-ray signature from Cyg X-3 was first claimed in \cite{Lamb77}, but since then there are many pieces of contradictory evidence in high-energy and very high-energy bands. Until recently, definite detections in the high-energy bands have been published by the AGILE (\citealt{Tavani09}) and \emph{Fermi} (\citealt{Abdo09}) collaborations, respectively. Furthermore, some more extended campaigns have been carried out by employing telescopes AGILE (\citealt{Bulgarelli12,Piano12}) and \emph{Fermi} LAT (\citealt{Williams11,Corbel12,Bodaghee13}). The significant features of the GeV $\gamma$-rays are given as follows: very high confidence, the correlated variability with the radio and X-ray emissions, and the strong orbital modulation at different epoches. Therefore, these detections are considered to be highly reliable.  Unfortunately, very high-energy $\gamma$-rays are still not detected by MAGIC (even during the activity epoches of the GeV $\gamma$-rays; \citealt{Aleksi10}) and VERITAS (\citealt{Archambault13}).

The GeV-band emissions from Cyg X-3 are associated with high-level soft X-ray and moderate radio emissions. Concretely, in order to detect the GeV emission, the following three conditions have to be satisfied (\citealt{Corbel12,Piano12}): (1) the soft X-ray count rate is above 3 counts $\rm s^{-1}$ in the 3--5 keV band; (2) the hard X-ray count rate is below 0.02 counts $\rm cm^{-2}\ s^{-1}$ in the $>15$ keV band; (3) the radio flux is above 0.2--0.4 $\rm Jy$ at 15 GHz. The detected GeV $\gamma$-rays have important unique features, whose physical properties have yet to be explored. First, the GeV emissions from Cyg X-3 is episodic rather than steady. Second, there exists a general characteristic for the GeV band observations, that is, they are detected each time when Cyg X-3 is moving into or out of the hypersoft X-ray (or ultrasoft X-ray) state, which corresponds to the quenched radio state. Third, a fact is that the published AGILE spectrum with index $2.0\pm 0.2$ is harder than the \emph{Fermi} LAT spectrum with index $2.7\pm 0.25$ (e.g., \citealt{Abdo09,Piano12}). Fourth, the modulation of the GeV band emission reaches 100 $\%$ in amplitude after background subtraction. This modulation is almost in anti-phase with X-rays, that is, the maximum flux of the GeV emission occurs at superior conjunction where it almost corresponds to X-ray minimum one.

From a theoretical point of view, \cite{Zdziarski10} have explained the complex X-ray energy spectra of Cyg X-3, assuming that the central compact object is surrounded by a thermal plasma cloud. Furthermore, the modulation of X-rays is interpreted as Thompson scattering of the X-rays when they pass through the strong stellar wind of the Wolf-Rayet star (\citealt{Zdziarski12b}), which requires an X-ray emission region to be close to the compact object.

In the high-energy bands, the orbital modulation of the GeV emission has been modeled in a jet model by \cite{Dubus10} and \cite{Zdziarski12a}. They concluded that the GeV band emission location is outside of the system separation, possibly up to 10 times orbital radii. On the other hand, the study regarding absorption of high-energy $\gamma$-rays infers that the GeV emissions should be at least located at distance $\sim10^{8}-10^{10}\rm \ cm$ from the central compact object (\citealt{Cerutti11}). Based on a steady-state jet model with stationary injection of high-energy electrons, \cite{Zdziarski12a} have modeled \emph{Fermi} LAT observations together with low-energy band data. The evident features are that both a low-energy break in the electron distribution and a relatively weak magnetic field are necessary. In the framework of the pair cascade model for microquasars, the \emph{Fermi} LAT spectral fitting is carried out in the inner jet (\citealt{Sitarek12}). It appears that the emission location is inconsistent with the results reported in \cite{Dubus10}, \cite{Cerutti11}, and \cite{Zdziarski12a}. By adopting the simplified leptonic and hadronic scenarios, \cite{Piano12} modeled the AGILE observations together with the hypersoft X-ray spectrum and MAGIC upper limits. Besides, the fittings to the AGILE observations have been carried out in a hadronic model (\citealt{Sahakyan14,Khiali14}).

Generally, an acceleration and/or emission region, in a persistent jet during the canonical LH state, continuously spans a large space range (\citealt{Romero03,Bosch06,Malz13,ZAA14,ZhangXL14}). In view of special properties of the HID of Cyg X-3 and its GeV band observations being associated with soft X-ray and moderate radio emissions, it seems that jet properties of Cyg X-3 are different from that in the LH state of other X-ray binaries. Furthermore, the GeV band emissions detected by AGILE and \emph{Fermi} LAT are over time-scales of days/weeks, which demonstrates that a continuous acceleration rather than an impulsive, single, adiabatic plasma ejection is at work (see also \citealt{Piano12}). Motivated by the GeV emission characteristics and studies reported in \cite{Corbel12} and \cite{Miller-Jones09}, we carry out a study of multi-waveband spectral energy distributions in a leptonic jet model, during the soft spectral state of Cyg X-3. Comparing with previous work (e.g., \citealt{Zdziarski12a}), we calculate the electron distribution including cooling processes via a kinetic equation (see Equation \ref{dNdz}). We find that the emission region in Cyg X-3 is rather compact and located at the jet height of about one orbital radius, which is similar to the scenario that all $\gamma$-ray emission models for blazars are one zone.

In the next section, we present a brief description of the model for Cyg X-3. Theoretical spectra confronting with multi-wavelength observations are described in Section 3. Section 4 contains the conclusions and discussion.

\section{Model Description}
In this study, the dissipation region is located at a certain location of the jet. The relativistic electrons accelerated in the jet dissipation region emit the non-thermal multi-wavelength emissions. Although there are many studies on internal shock interactions producing dissipation in jets from both X-ray binaries and active galactic nuclei (e.g., \citealt{Spada01,Jamil10}), the physical process of how relativistic electrons are accelerated remains unclear. Besides the internal shock acceleration, the reconfinement shock (\citealt{Dubus10,Zdziarski12a}) and the magnetic reconnection (e.g., \citealt{Lyubarsky01,Lyubarsky05,Lyubarsky10,Kagan13,Sironi14}) seem to be possible as well. In this study, we focus on the broadband emission spectra, but do not study the dissipative process in the plasma material of the jet. The geometry of the model is similar to Figure 1 of \cite{ZhangXL14}, but the emission region is more compact than that of Cyg X-1.

The steady-state electron distribution in a conical jet has been explicitly clarified in the recent literature (\citealt{ZAA14} and refs. therein). Following this work, the equation for relativistic electrons in the dissipation region is written as
\begin{equation}
\frac{1}{z^2}{\partial \over \partial z} [\Gamma \beta_{\Gamma}cz^2 N(\gamma,z)] + {\partial \over \partial
\gamma} \left[\Gamma \beta_{\Gamma}cN(\gamma,z) {d\gamma \over  dz}\right] = Q, \label{dNdz}
\end{equation}
where $\Gamma$ is the bulk Lorentz factor of the dissipation region, $\beta_{\Gamma} = \sqrt{\Gamma^2-1}/\Gamma$ the bulk velocity, $\gamma$ the electron Lorentz factor, and $c$ the speed of light. $ N(\gamma,z)$ stands for electron number density per unit volume, as a function of the electron energy $\gamma$ and the jet height $z$. Here, $z$ is the distance with respect to the central compact object. The energy change of the accelerated electrons along the jet is given as
\begin{equation}
{d\gamma \over dz} = {1 \over  c\beta_{\Gamma} \Gamma}\left(d\gamma \over dt^{'}\right)_{\rm rad}- {2 \over
3}{\gamma \over z}, \label{dgdz}
\end{equation}
where $dt^{'}$ is the proper time. The factor 2/3 indicates a two-dimensional adiabatic expansion of the dissipation region. The total radiative loss rates of an electron, $(d\gamma/dt^{'})_{\rm rad}$, include synchrotron emission, self-Compton scattering, external Compton scattering of the photons from the companion and disk.

Using the same approach as given in \cite{ZAA14}, one can change Equation (\ref{dNdz}) into the form of Equation (27) of \cite{ZAA14}, which has the similar form as corresponding equations in \cite{Sikora01} and \cite{Moderski03}.\footnote{\cite{ZhangXL14} have considered the escape loss of electrons, which is not necessary. Because the motion of the electrons upward has already been taken into account by the term $\partial N(\gamma,z)/\partial z$, but the electrons are confined by magnetic field and do not escape sideways. In the aspect of numerical calculations, the added escape term only has a slight influence on the spectra of the electrons with $\gamma < 100$, nevertheless, the total emission spectra presented in the paper are not affected.} Furthermore, the accelerated electron is injected at the dissipation region between $z$ and $2z$ with a broken power-law form
\begin{equation}
Q (\gamma)=K \frac{1}{\gamma^{p1}\gamma_{\rm br}^{p2-p1}+\gamma^{p2}},
\end{equation}
where $\gamma_{\rm br}$ is the break energy of the relativistic electron, $p1$ is the spectral index of electrons below $\gamma_{\rm br}$, and $p2$ is the electron spectral index above $\gamma_{\rm br}$. The normalization constant of the electrons, $K$, is determined by
\begin{equation}
{L_{\rm rel}}=Km_{\rm e}c^2\int_{V}dV\int^{\gamma_{\rm max}}_{\gamma_{\rm min}}\frac{1}{\gamma^{p1}\gamma_{\rm br}^{p2-p1}+\gamma^{p2}}\gamma d\gamma, \label{Lrel}
\end{equation}
where $V$ is the volume of the dissipation region. $\gamma_{\rm max}$ and $\gamma_{\rm min}$ are the maximum and minimum energies of the accelerated electron, respectively. $\gamma_{\rm max}$ can be obtained by balancing the acceleration and cooling rates. In this work, we use the shock acceleration mechanism to obtain maximum energy of electrons as that in \cite{ZhangXL14}. The acceleration efficiency $\eta$ and $\gamma_{\rm min}$ is set as 0.1 and 1, respectively. A fraction of the jet power, $q_{\rm rel}=L_{\rm rel}/L_{\rm jet}$, is convected to accelerate electrons, in which $L_{\rm jet}$ is assumed to be proportional to the accretion power $L_{\rm acc}=\dot{M} c^2$, e.g., $L_{\rm jet}=q_{\rm jet}L_{\rm acc}$. Here, $\dot{M}$ is the mass accretion rate, $q_{\rm rel}$ is a free parameter, and $q_{\rm jet}$ is 0.5 for simplification.

In this model, except the photon field of an accretion disk, which is in the form of the multi-temperature blackbody spectrum (e.g., \citealt{Kato08}), all the radiation formulae used are similar to those presented in \cite{ZhangXL14} for the LH state of Cyg X-1. We thus skip these fundamental descriptions; interested readers are referred to that work.

\section{Modeling Spectral Energy Distributions}
Cyg X--3 is a well-known compact X-ray binary, which is composed of a black hole or neutron star and a Wolf-Rayet companion, with an orbital period of 4.8 hours, at a distance $\sim 7.2\ \rm kpc$ (\citealt{Ling09}). The nature of the central compact object remains unclear. Although a recent work is inclined to support the existence of a low-mass black hole (\citealt{Zdziarski13}), we still use a large value $\sim 20\ M_{\odot}$ in this study (\citealt{ Cherepashchuk94}), which was also adopted in \cite{Dubus10} and \cite{Zdziarski12a}. The other parameters of the system are following \cite{ Cherepashchuk94}: the radius of companion star is $R_{\star}=2\times10^{11}\ \rm cm$; its surface temperature is $T=9\times10^{4}\ \rm K$; and the orbital radius is $d = 4.13\times10^{11}\rm \ cm$, which is deduced from the Kepler's law. Furthermore, the parameters associated with the jet are the bulk velocity $\sim0.81c$, the viewing angle $14^{\circ}$ (\citealt{Mioduszewski01}), and the half-opening angle $\sim 5^{\circ}$ (\citealt{Miller-Jones06}).

As mentioned in Section 1, a great number of observational investigations on Cyg X-3 have been carried out since 1960s. But so far, the simultaneous multi-waveband data have not been  obtained yet. In particular, although GeV band observations have been firmly confirmed (\citealt{Abdo09,Tavani09}), their spectral shapes remain uncertain due to the episodic nature of the $\gamma$-ray emission. The \emph{Fermi }LAT average spectral index $2.70\pm0.25$ (between 100 MeV and 100 GeV) is derived by using the accumulated data for about four months (\citealt{Abdo09}). However, by integrating the peak $\gamma$-ray events observed by AGILE, the average differential spectral index (between 50 MeV and 3 GeV) is fitted in $2.0\pm0.2$ (\citealt{Piano12}).  The spectral differences between them were explained as the scenario that there is a fast hardening in the spectra during the main $\gamma$-ray events (\citealt{Piano12}). In this work, we use the AGILE and \emph{Fermi} LAT data together with radio, IR, and X-ray data (corresponding to the canonical HS state), as well as VERITAS upper limits (lower than MAGIC upper limits reported in \citealt{Aleksi10}) to limit our theoretical spectra.

\begin{figure*}[]
  \begin{center}
  \begin{tabular}{ccc}
\hspace{-0.79cm}
     \includegraphics[width=80mm,height=70mm,bb=20 260 455 600]{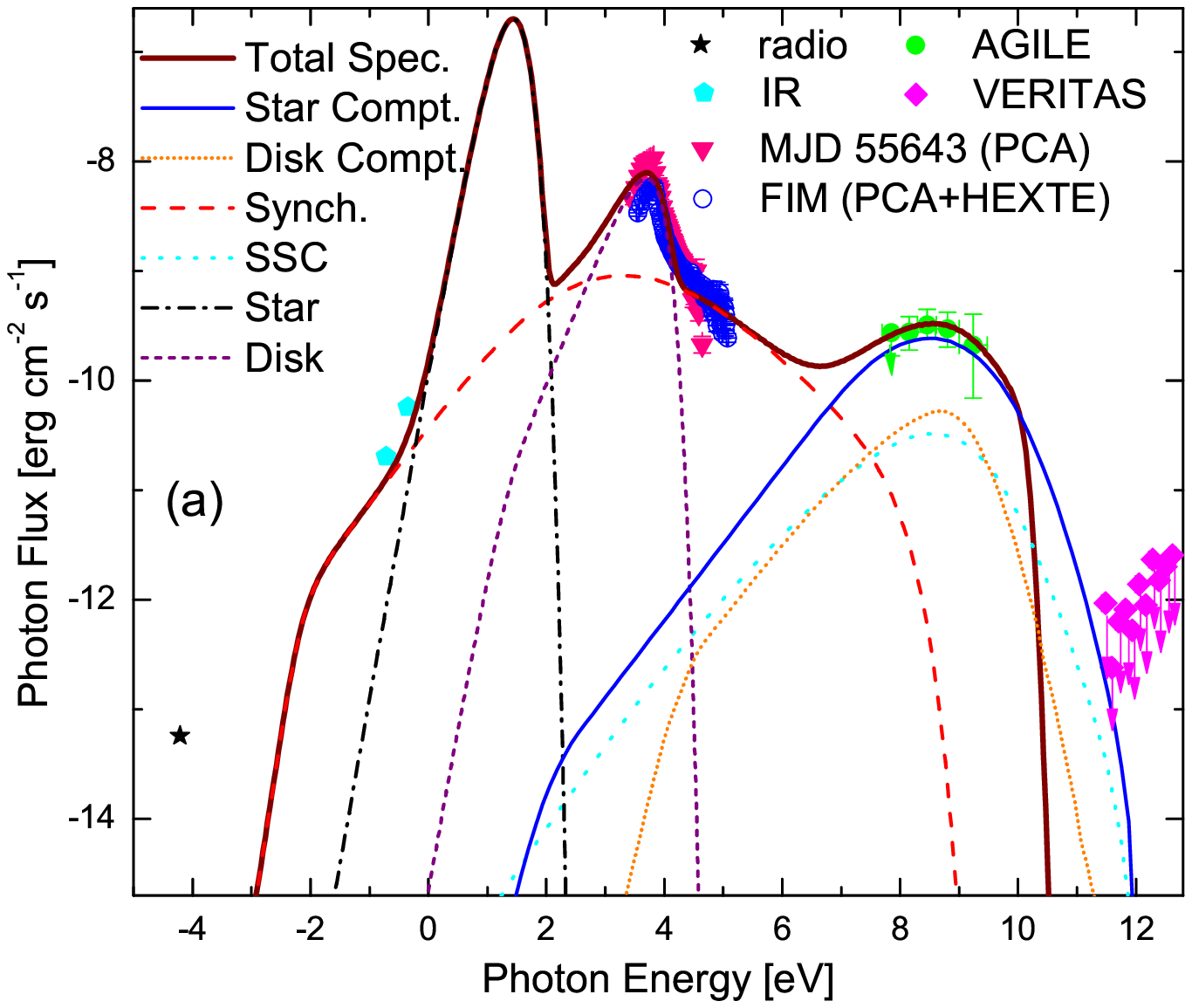}& \ \ \
\hspace{-0.79cm}
     \includegraphics[width=80mm,height=70mm,bb=20 260 455 600]{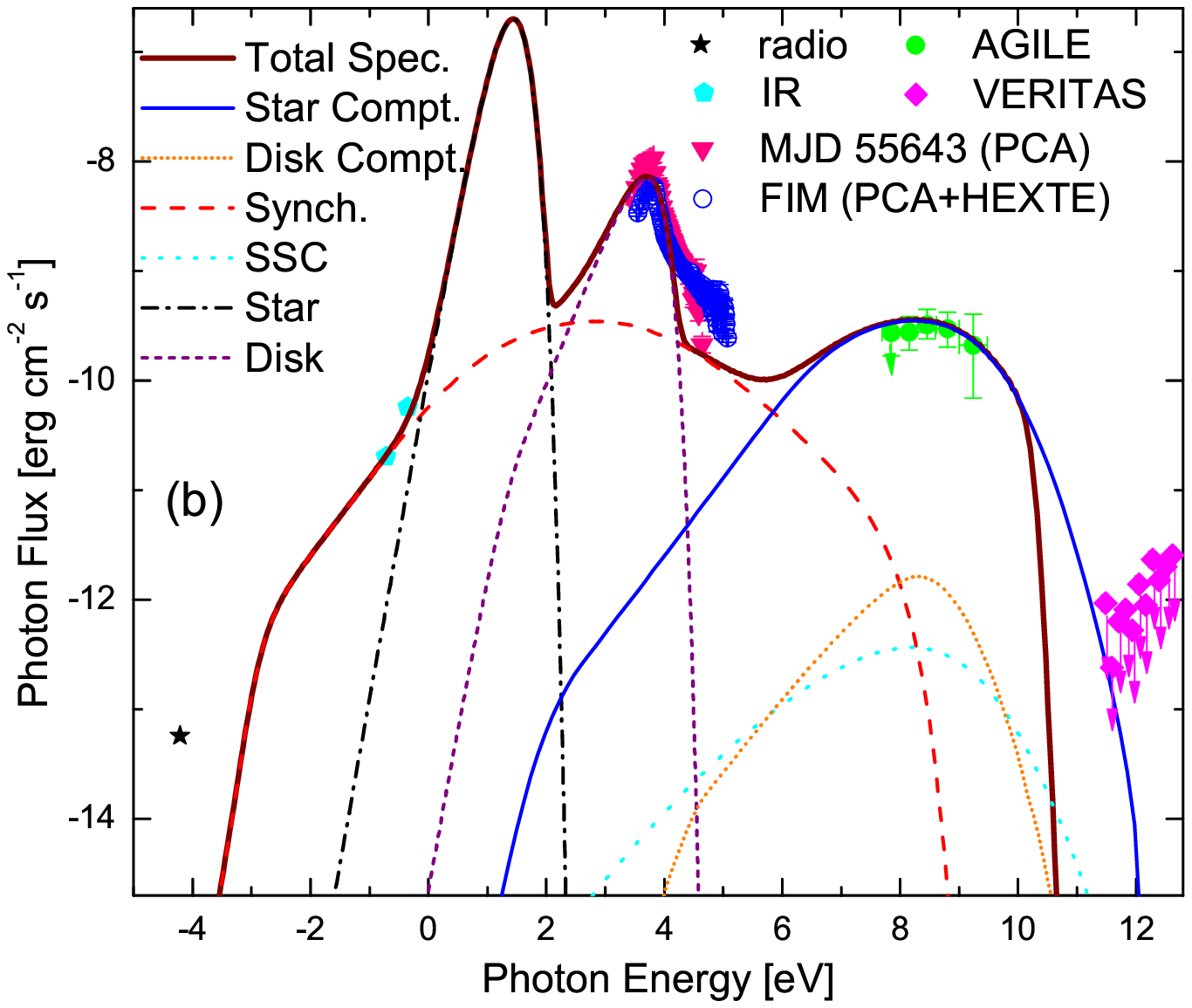}\\
\hspace{-0.79cm}
     \includegraphics[width=80mm,height=70mm,bb=20 260 455 600]{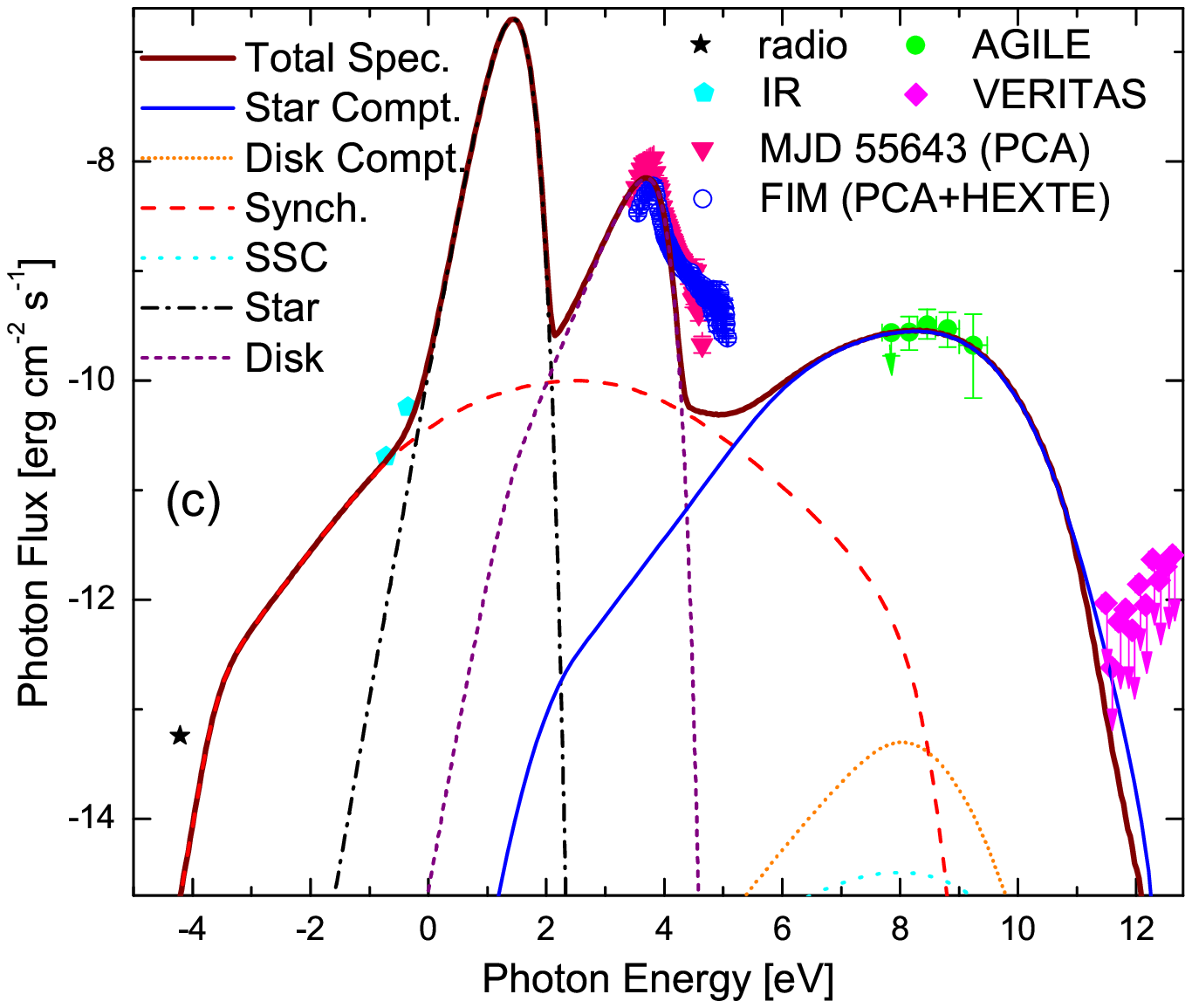}&\ \ \
\hspace{-0.79cm}
     \includegraphics[width=80mm,height=70mm,bb=20 260 455 600]{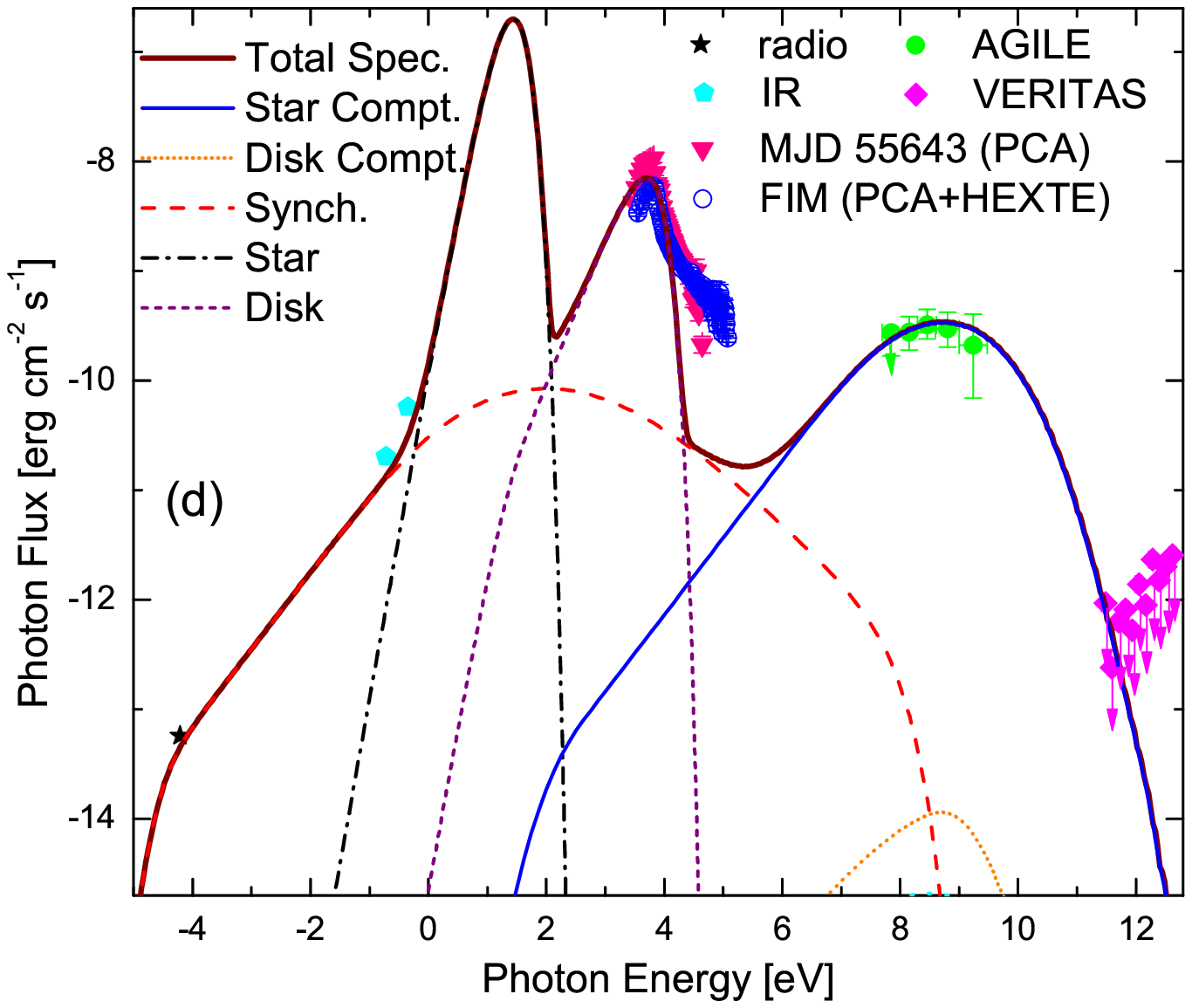}& \ \ \
\end{tabular}
  \end{center}
\caption{Multi-waveband emission spectra of Cyg X-3 during the peak $\gamma$-ray events. The fitting parameters in panels (a)-(d) are listed in Table 1 for Case A1--A4, respectively. The broadband observations are radio data from \cite{Abdo09} and \cite{Zdziarski12a}, IR data from \cite{OBF01} and \cite{Zdziarski12a}, average X-ray data in the FIM state (by PCA+HEXTE) from \cite{Koljonen10}, X-ray data on JMD 55643.0 (by PCA) from
\cite{Corbel12}, AGILE data from \cite{Piano12}, and VERITAS upper limits from \cite{Archambault13}. The total energy spectra (thick solid line) consider the attenuation of high-energy photons by absorption of the companion photons.}  \label{figs:AGILE}
\end{figure*}

\begin{figure*}[]
  \begin{center}
  \begin{tabular}{ccc}
\hspace{-0.79cm}
     \includegraphics[width=80mm,height=70mm,bb=20 260 455 600]{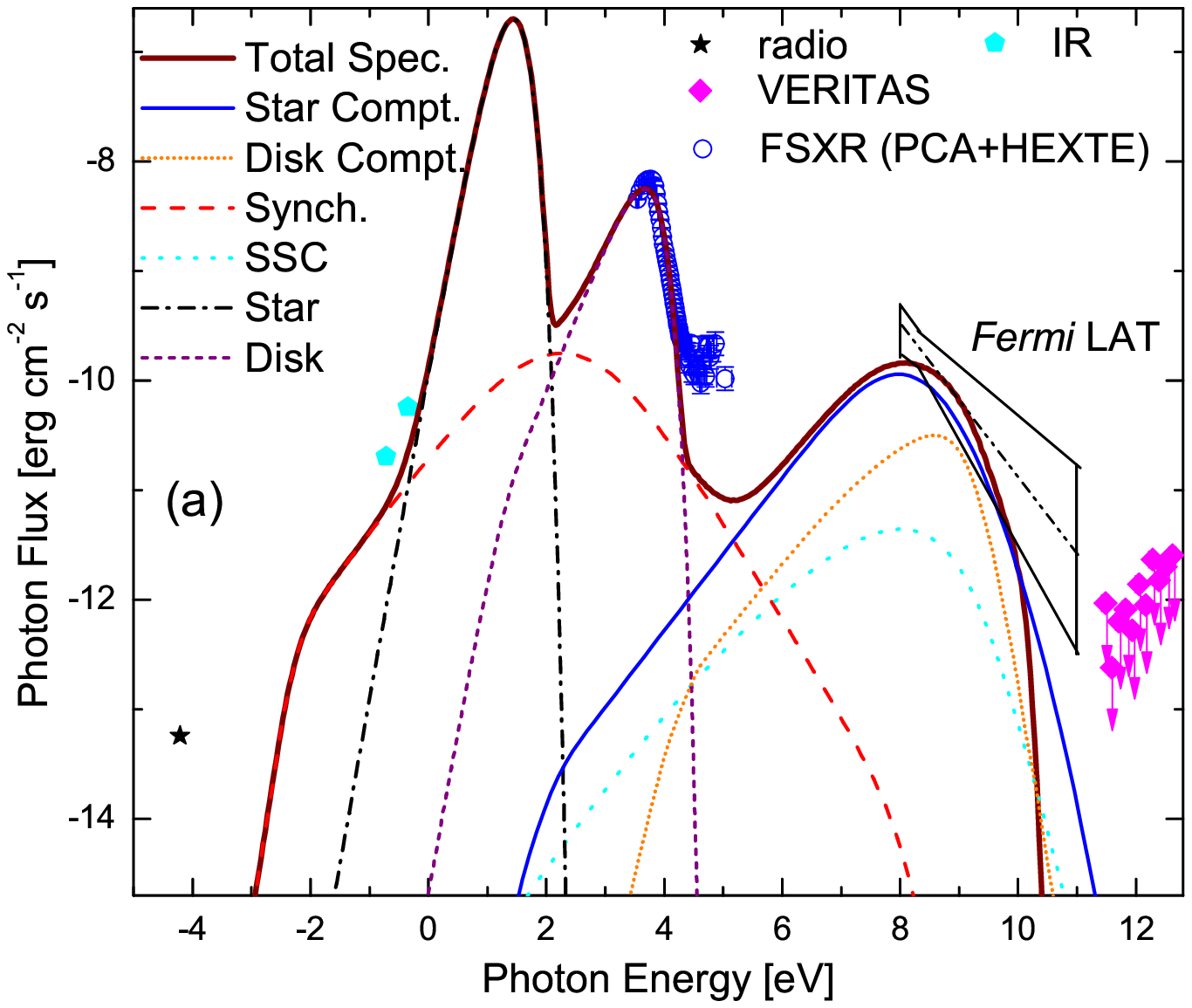}& \ \ \
\hspace{-0.79cm}
     \includegraphics[width=80mm,height=70mm,bb=20 260 455 600]{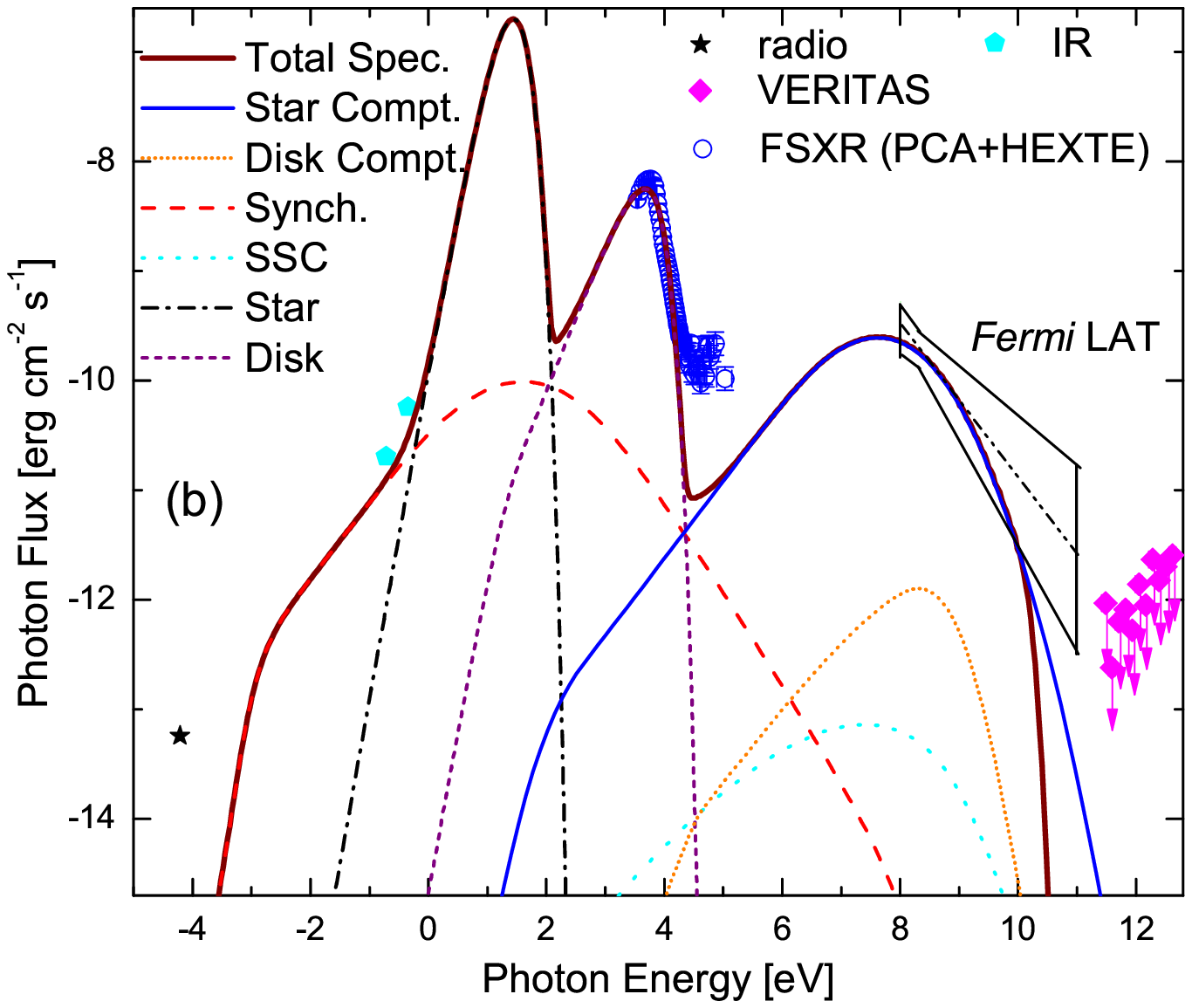}\\
\hspace{-0.79cm}
     \includegraphics[width=80mm,height=70mm,bb=20 260 455 600]{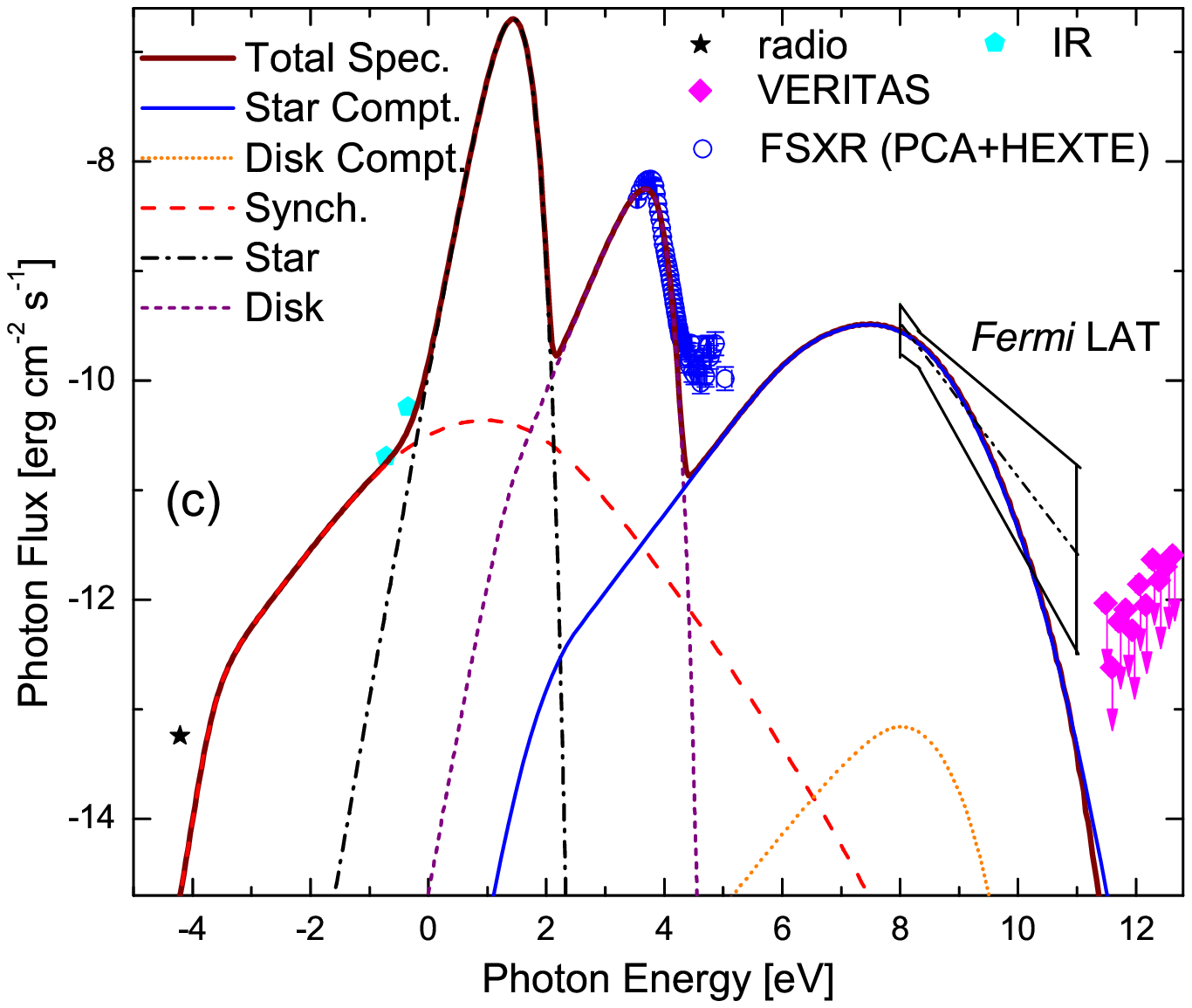}&\ \ \
\hspace{-0.79cm}
     \includegraphics[width=80mm,height=70mm,bb=20 260 455 600]{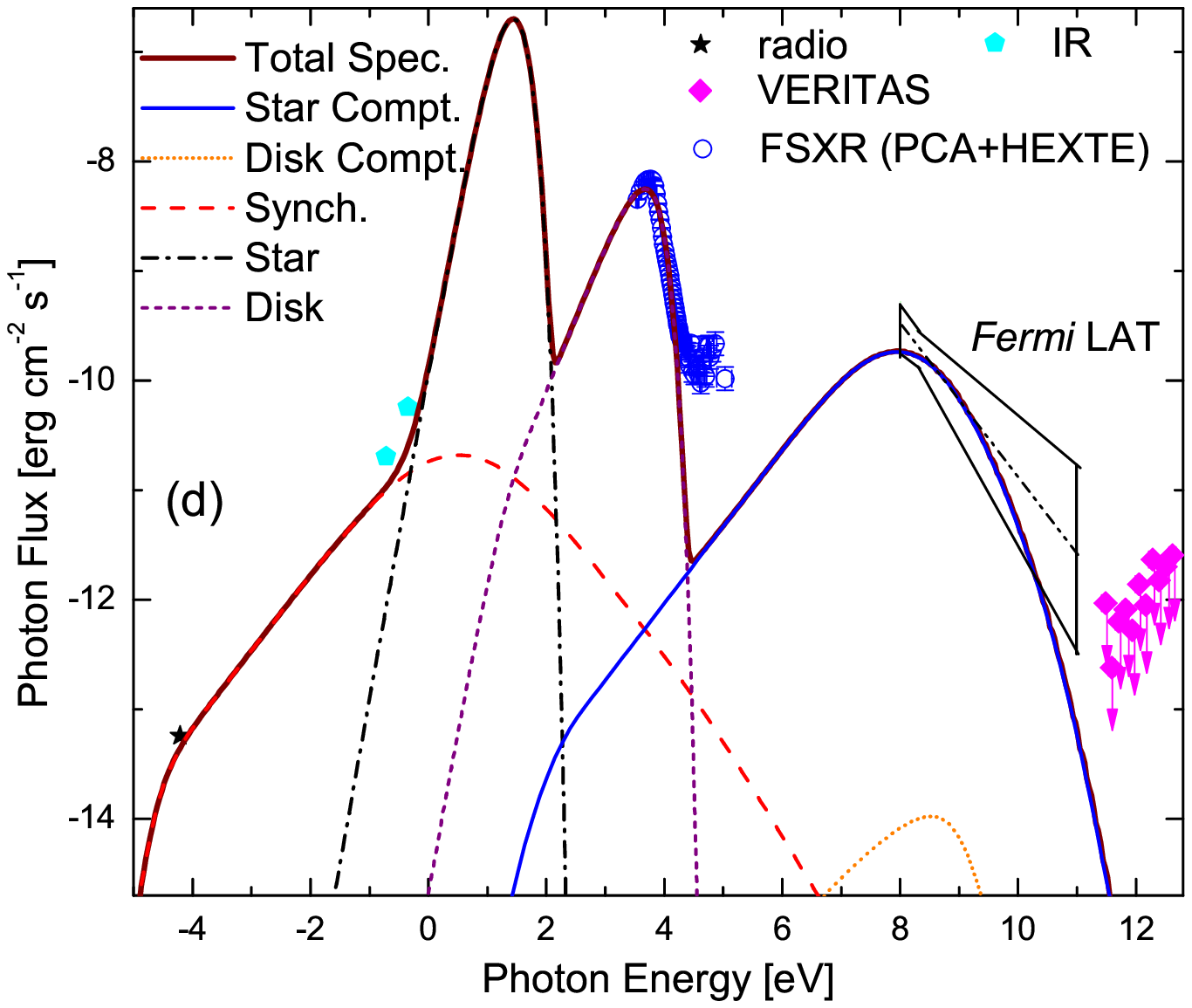}& \ \ \
\end{tabular}
  \end{center}
\caption{Broadband spectral energy distributions of Cyg X-3 compared to the average \emph{Fermi} LAT observations. The parameters used in panels (a)-(d) are listed in Table 1 for Case B1--B4, respectively. The observations are radio data from \cite{Abdo09} and \cite{Zdziarski12a}, IR data from \cite{OBF01} and \cite{Zdziarski12a}, average X-ray data in the FSXR state (by PCA+HEXTE) from \cite{Koljonen10}, and VERITAS upper limits from \cite{Archambault13}. The error contour and dot-dot-dashed line indicate an average power-law fit of the \emph{Fermi} LAT observations, with spectral index 2.70 $\pm$ 0.25, integrating the two active windows for about four months. The total energy spectra (thick solid line) include the attenuation of high-energy photons by the companion photons.}  \label{figs:Fermi}
\end{figure*}

The average radio flux during the $\gamma$-ray active periods is about 0.38 $\rm Jy$ at 15 GHz (\citealt{Abdo09,Zdziarski12a}). However, it is unclear whether the radio synchrotron spectrum related to the $\gamma$-ray activity is optically thin or thick (\citealt{Corbel12}). In the X-ray and high-energy ranges, the multiple sets of RXTE/PCA spectra, being associated with the $\gamma$-ray flaring process, are reported in \cite{Corbel12} during the soft to hard state transition. In particular, X-ray spectral distributions observed on MJD 55643.0 (\citealt{Corbel12}), which are very close to the FIM spectra reported in \cite{Koljonen10}, corresponds to the peak of the $\gamma$-ray flare. Furthermore, the AGILE data used in this study have been analyzed by employing the peak $\gamma$-ray flare events. In the first set of fitting (i.e., Case A1--A4), we thus adopt the AGILE data together with the FIM data, RXTE/PCA data on MJD 55643.0, radio and IR data, and VERITAS upper limits (see those plotted on Figure \ref{figs:AGILE}). In the second set of fitting (i.e., Case B1--B4), we adopt the \emph{Fermi} LAT observations, radio, IR, and FSXR data, as well as VERITAS upper limits (see Figure \ref{figs:Fermi}). We note that the hypersoft and ultrasoft data are used in \cite{Piano12} and \cite{Sitarek12}, respectively. It appears that during these states, in particular, the hypersoft state, there is no jet production (\citealt{Koljonen10}). It should be emphasized that the FIM and FSXR states correspond to the canonical HS state (\citealt{Koljonen10}).

\begin{deluxetable}{cccccccc}
\tabletypesize{}
\tablecaption{Fitting Parameters for the Emission Spectra of Cyg X-3.}
\tablewidth{0pt}
\tablehead{
\colhead{Case} & \colhead{$\dot{M}(M_{\rm \odot}\ $yr$^{-1})$} & \colhead{$z$} &
\colhead{$\gamma_{\rm br}$} & \colhead{$q_{\rm rel}$} & \colhead{$B'_{\rm 0}(\rm G)$} & \colhead{$p1$} & \colhead{$p2$}}
\startdata
A1 & $2.0\times10^{-8}$ & $0.01 d$ & $8.0\times10^{3}$  & 0.44 & $5.4\times 10^{3}$ & 1.6 & 3.0\\
A2 & $2.0\times10^{-8}$ & $0.1 d$ & $8.0\times10^{3}$  & 0.19 & $2.8\times 10^{3}$ & 1.6 & 3.0\\
A3 & $2.0\times10^{-8}$ & $1 d$ & $8.0\times10^{3}$  & 0.10 & $1.0\times 10^{3}$ & 1.6 & 3.0\\
A4 & $2.0\times10^{-8}$ & $10 d$ & $8.0\times10^{3}$  & 0.21 & $9.8\times 10^{1}$ & 1.6 & 3.0\\
\tableline
B1 & $1.6\times10^{-8}$ & $0.01 d$ & $2.0\times10^{3}$  & 0.47 & $4.0\times 10^{3}$ & 1.6 & 3.8\\
B2 & $1.6\times10^{-8}$ & $0.1d$ & $2.0\times10^{3}$  & 0.25 & $2.0\times 10^{2}$ & 1.6 & 3.8\\
B3 & $1.6\times10^{-8}$ & $1 d$ & $2.0\times10^{3}$  & 0.22 & $7.0\times 10^{2}$ & 1.6 & 3.8\\
B4 & $1.6\times10^{-8}$ & $10 d$ & $2.0\times10^{3}$  & 0.35 & $8.0\times 10^{1}$ & 1.6 &  3.8
\enddata
\tablenotetext{*}{Note. Symbol indicating $\dot{M}$: accretion rate; $z$: location of dissipation region in jet; $\gamma_{\rm br}$: break energy of electron; $d$: separation of system; $q_{\rm rel}$: conversion efficiency; $B'_{\rm 0}$: magnetic field strength; $p1$: spectral index of electron below $\gamma_{\rm br}$; $p2$: spectral index of electron above $\gamma_{\rm br}$.}
\label{table:cases}
\end{deluxetable}

The results of the first set of fitting are presented in Figure \ref{figs:AGILE}. The non-thermal spectral energy distributions include synchrotron emission, self-Compton scattering, anisotropic inverse Compton scattering of the photons from the companion and disk. Meanwhile, thermal spectra, that is, the blackbody spectrum of the Wolf-Rayet companion and the multi-temperature blackbody spectrum of the accretion disk, are also plotted. Assuming that the emission regions are located at different heights of the jet, $z=0.01d$, $0.1d$, $1d$, and $10d$ ($d$ is the orbital separation of the system), we model the multi-waveband spectral energy distributions. The parameters used are listed in Table \ref{table:cases} for Case A1--A4, which correspond to panels (a)--(d), respectively. In order to exclude the influence of the diversity of possible for electron spectral distributions on emission spectra, we fix the parameters $\gamma_{\rm br}$, $p1$ and $p2$ in each set of fittings. The summed total spectra have considered the attenuation of $\gamma$--$\gamma$ interactions due to the companion photons at the superior conjunction.

The four scenarios that we fit in Figure \ref{figs:AGILE} are elucidated as follows. (1) As shown in panel (a), the synchrotron emission process can produce fluxes at IR and X-ray bands, and the AGILE observations can be well fitted by the sum of synchrotron self-Compton (SSC) and inverse Compton scattering of the photons from the companion and disk. At about 0.1 TeV, there is an evident absorption due to $\gamma$--$\gamma$ interactions. (2) In panel (b), the synchrotron spectrum can reproduce the IR observation. But SSC and disk photon Comptonization become weaker in this case. Even through the inverse Compton scattering of companion photons can roughly fit the AGILE data, the theoretical spectrum has over-produced emissions at the low-energy part of the AGILE data. Similar to that of panel (a), $\gamma$--$\gamma$ absorptions are also significant. (3) For panel (c), IR emissions can be fitted by the synchrotron emission. The AGILE data and VERITAS upper limits can be matched by the inverse Compton scattering of the photons of the companion. There also exists a slight attenuation at very high-energy bands due to pair production. (4) In panel (d), the synchrotron emission can produce the emission fluxes ranging from radio to IR bands. The inverse Compton scatting of companion photons can predict the TeV band emission, but does not fit the high-energy part of the AGILE observations. SSC component is negligible and is not shown in this panel. In addition, $\gamma$--$\gamma$ absorptions are neglected completely in this case.

By comparing the above fittings, we find that the results of the fitting in panel (c) is most likely scenario among them. Because the theoretical spectra can well reproduce observations at both IR and GeV bands, and also can predict the TeV-band observations. The further reason is given as follows. The fitting results in panel (c) conform to the required conditions (see Section 1) in order to detect the GeV $\gamma$-ray emissions. Besides, the 100 per cent orbital modulation implies that the emissions at GeV energy are from anisotropic inverse Compton scattering of the companion photons as opposed to that of panel (a), which has contributions from both SSC and Comptonization of the companion and disk photons.

We turn now to investigate spectral energy distributions by using the \emph{Fermi} LAT data together with the other waveband data. The fitting results are presented in Figure \ref{figs:Fermi}. The used parameters are listed in Table \ref{table:cases} for Case B1--B4, which correspond to panels (a)--(d), respectively. Given that some basic descriptions are similar to the scenario in Figure \ref{figs:AGILE}, we would not repeat here it (see also the caption of Figure \ref{figs:Fermi}). In this case, we adopt a lower-energy break, $\gamma_{\rm br}=2\times10^3$, and a softer electron spectral index 3.8 above $\gamma_{\rm br}$, comparing with the first set of fittings. As shown in panels (a)-(d), the total spectra can produce the TeV band flux, but the flux becomes lower due to more rapidly cooling of high-energy electrons. We could exclude panels (a) and (b), because of the current observational constraint, that is, the very strong orbital modulation at GeV energy bands. The best fitting is that in panel (c), but due to the uncertainties in the \emph{Fermi} LAT spectral shape, we cannot completely exclude the possibility in panel (d).

In summary, we obtain the fittings to the \emph{Fermi} LAT and AGILE data together with the other band data and the VERITAS upper limits, assuming that dissipation regions are located at different heights of the jet. From these two set fittings, we find that the currently observed emission at the GeV band should originate from a rather compact dissipation region at the height about $1d$. The GeV emissions are produced by an anisotropic inverse Compton scattering of the photons of the companion, which is consistent with the results arising from orbital modulations (\citealt{Abdo09,Dubus10,Zdziarski12a}) and constraints of pair production due to internal absorption with the ambient X-rays (\citealt{Cerutti11}). Furthermore, the multi-temperature blackbody spectrum of the disk can explain soft X-ray emissions and the synchrotron emission can reproduce observations at the IR band. The occurrence of a moderate radio emission, which is required to detect the GeV signature, is not in the same zone but should be from the distance $>1d$. This model could be tested by the polarization measurement at IR bands. A correlation study between IR and GeV bands is also possible to test our results. The GeV emission should be related to the TeV band emission that originates from the scale of the binary system, which is detectable by the upcoming Cherenkov Telescope Array.

We infer that there exits a single dissipation region at the distance about $1 d$ during the period of $\gamma$-ray activities, which dissipates its energy to accelerate electrons that produce the multi-band emissions. During the $\gamma$-ray activity, if there exits the other dissipation regions at smaller distance than this location, they will produce an observable signature and reduce the net modulation amplitude. The observational characteristics that the radio outburst is delayed by days with respect to $\gamma$-rays (e.g., \citealt{Abdo09}), could be explained as separate dissipation regions being formed at much larger heights of the jet, which may be due to very different dissipative processes (see also \citealt{Zdziarski12a}). But at these large heights, it is more difficult to detect $\gamma$-ray emissions due to a decreasing in soft photon density of the companion. Generally, radiative losses of high-energy electrons are so rapid that they are immediately radiated at the accelerated location. However, low-energy electrons accelerated at a small height of the jet, which have a longer cooling time scale, could escape to a large height of the jet to produce emissions and enhance radio emission fluxes. Therefore, an outburst activity in a large scale could be occasionally accompanied by the dissipative processes before occurring at a small scale.

\section{Conclusions and Discussion}
We have proposed a radiation model to investigate the multi-band emission from the Galactic X-ray binary Cyg X-3, during the canonical HS state. Considering that the dissipation region is confined in the jet, we calculate broadband spectral energy distributions to confront with the AGILE and\emph{ Fermi} LAT data together with the other band data during two soft X-ray spectral states. The results demonstrate that our model can explain the current multi-frequency observations and predict the underlying TeV emission. Furthermore, a dissipation region being not extended can be confined at the distance about $1 d$, to accelerate electrons that reproduce multi-waveband observations. We note that virtually all gamma ray emission models for blazars are one zone, and thus are similar to that for Cyg X-3. The synchrotron process is responsible for the emissions at the IR band, and the GeV and TeV band emissions are from the inverse Compton scattering of the companion photons. The multi-temperature blackbody emission reproduces the observation at the soft X-rays, which exactly corresponds to the canonical HS state (i.e., high accretion rate and standard thin disk).

The present work attempts to understand the main characteristics of multi-band observations from the X-ray binary Cyg X-3. The GeV emissions prior to and after the hypersoft state should be due to the presence of dissipation process at the location about $1 d$ when the jet turns off/on. The strong orbital modulation at the GeV band implies that the dissipation region is approximately stationary, or else the emissions that are produced at different heights would dilute the modulation amplitude. A few days' delay between the onset of the GeV emission and major radio flare has been explained as a propagation effect of relativistic ejecta (e.g., \citealt{Abdo09}). We suggest that the strong dissipation process, which is formed at later days at large scale of the jet, dissipates their energy to produce a major radio outburst, although it is possible that the low-energy electrons escape to the large scale to produce stronger radio emission.

Even if the soft X-ray spectral state has been confirmed to associate with the GeV-band emissions (e.g., \citealt{Abdo09}), the detailed spectral connection between the soft X-ray state and the GeV band is yet unclear. In the current work, we have used the flaring states (e.g., FIM and FSXR) of Cyg X-3 defined in \cite{Koljonen10} to correspond to the AGILE and \emph{Fermi} LAT detections. Such an approach may be slightly arbitrary, but our results are not affected since the high-energy tail of X-ray emissions is generally from the disk/corona region (e.g., \citealt{Zdziarski12b}). Although panel (a) of Figure \ref{figs:AGILE} presents the fitting to X-ray high-energy tails, it has been excluded due to the strong orbital modulation at GeV bands.

In this study, we have used distributions of injection electrons with a broken power-law form, which has been commonly adopted in studies of blazars observed by \emph{Fermi} LAT (e.g., \citealt{Ghisellini11}). The break energy $\gamma_{\rm br}$ used in this work corresponds to the anticipated threshold of a diffusive shock acceleration (e.g., \citealt{Stawarz07}). On the other hard, the high accretion rates are adopted in order to fit the RXTE/PCA data, which exactly corresponds to a standard thin accretion disk mode.

It is an open issue with regard to the question of whether jets are launched in the soft state, that is, in the thin disk mode (see also \citealt{Russell11}). The leading model for the jet formation mechanism is the Blandford-Znajek and Blandford-Payne mechanisms (\citealt{Blandford77,Blandford82}), which tends to generate a continuous jet unless there exists high instability in the disk, and requires the presence of a large-scale open magnetic field. For the episodic jet production mechanism,  \cite{Yuan09} initially suggested a magnetohydrodynamical model by analogy with the coronal mass ejections in the Sun. A general view is that thin-disk flows do not have strong large-scale magnetic field, therefore, which should not produce strong jets (\citealt{Meier01}). Whereas the large-scale magnetic field could be also produced effectively in the case of a thin disk, when the radial velocity of the accretion disk significantly increases due to the presence of the outflows (\citealt{Cao13}).

Furthermore, the standard disk is thought to be more suitable for the formation of powerful jets in the magnetized accretion ejection structure (e.g., \citealt{Ferreira97}), because the radial magnetic tension overcomes the toroidal one at the disk surface when the disk becomes too thick. \cite{Combet06} have advanced these works to propose a jet-emitting disk model, explaining the canonical spectral states of black hole X-ray binaries. In this framework, spectral energy distributions of the jet-emitting disk model have been investigated (\citealt{ZhangX13}).

\acknowledgments We thank the anonymous referee for constructive criticism and suggestions that significantly improved our manuscript. We thank Dr Wei-Min Gu, Dr Da-Bin Lin for critical reading of the manuscript, and Dr Mai-Chang Lei for useful discussions. We also appreciate Dr St\'{e}phane Corbel, Dr Karri Kolojonen, and Dr Giovanni Piano for providing us with the observational data. This work is partially supported by the National Natural Science Foundation of China (grant nos. 11233006 and 11363003).

\end{document}